\documentclass[conference]{IEEEtran}

\usepackage{amsmath,amsfonts}
\usepackage{cite}

\usepackage[dvips, final]{graphicx}
\graphicspath{{./eps/}} \DeclareGraphicsExtensions{.eps}
\newcommand{\figwidth}{.9\linewidth}

\newtheorem{theorem}{Theorem} \newcounter{remark}
\def\remark{\addtocounter{remark}1\noindent\emph{Remark \arabic{remark}:} }
\newcounter{testcase}
\def\testcase{\addtocounter{testcase}1\smallskip %
\noindent\emph{Test Case \arabic{testcase}:} }

\begin{document}
\title{A Novel Uplink Data Transmission Scheme For Small Packets In Massive MIMO System}

\author{\IEEEauthorblockN{Ronggui Xie$^1$, Huarui Yin$^1$, Zhengdao Wang$^2$, and Xiaohui Chen$^1$}
\IEEEauthorblockA{$^1$Department of Electronic Engineering and Information
Science, University of Science and Technology of China\\ \small
$^2$Department of Electrical and Computer Engineering, Iowa State
University }

}

\maketitle

\begin{abstract}
Intelligent terminals often produce a large number of data packets of small
lengths. For these packets, it is inefficient to follow the conventional
medium access control (MAC) protocols because they lead to poor utilization
of service resources. We propose a novel multiple access scheme that
targets massive multiple-input multiple-output (MIMO) systems based on
compressive sensing (CS). We employ block precoding in the time domain to
enable the simultaneous transmissions of many users, which could be even
more than the number of receive antennas at the base station. We develop a
block-sparse system model and adopt the block orthogonal matching pursuit
(BOMP) algorithm to recover the transmitted signals. Conditions for data
recovery guarantees are identified and numerical results demonstrate that
our scheme is efficient for uplink small packet transmission.
\end{abstract}

\IEEEpeerreviewmaketitle

\section{Introduction}

As intelligent terminals such as smart phones and tablets get more popular,
they produce an increasing number of data packets of short lengths. Modern
mobile applications that produce such small packets include instant
messaging, social networking, and other services \cite{re2},\cite{re3}.
Although the lengths of messages are relatively short, small packet
services put great burden on the communication network. Two kinds of
messages contribute to the traffic of small packets: one is the small
packets of conversation produced by active users that occupy only a small
percentage of the total online users \cite{re3}; the other is the signaling
overheads needed to transmit these conversation packets \cite{re4}.

In current wireless communication systems, a user follows the medium access
control (MAC) protocols to obtain the service resources. Either resources
are preallocated to the users in a noncompetitive fashion, or certain
random access scheme with collision resolution is used. For small and
random packets, the reservation-based approach is inefficient in resource
utilization due to irregularity of the packets. The collision-resolution
based approaches, on the other hand, can suffer from too many
retransmissions due to frequent collisions.

Recently, massive multiple-input multiple-output (MIMO) was studied as a
way to improve the system throughput of cellular systems
\cite{re9}-\cite{re12}. In massive MIMO systems, the number of antennas at
the base station (BS) can be more than the number of active single-antenna
users that are simultaneously served. When the number of antennas at BS is
large, the different propagation links from the users to the BS tend to be
orthogonal, and the large amount of spatial degrees of freedom are useful
for mitigating the effect of fast fading \cite{re10},\cite{re11}. Overall,
massive MIMO technique provides higher data rate, better spectral and
energy efficiencies \cite{re12}. All these advantages make massive MIMO a
promising technique.

In this paper, we propose a novel uplink small packet transmission scheme
based on precoding at the transmitters and sparsity-aware detection at the
receiver. The main motivation is to allow for a large number of users to
transmit simultaneously, although each user may be transmitting only a
small amount of data. Besides frame-level synchronization, no competition
for resources or other coordinations are required. This saves the signaling
overhead for collision resolution, and improves the resource utilization
efficiency.

The contributions of our work are as follows:
\begin{enumerate}

\item \emph{block-sparse system model is established:} We apply block
precoding at each transmitter in time domain, and by considering the user
activities, develop a block-sparse system model \cite{re20}-\cite{re25}.

\item \emph{conditions for signal recovery are given:} The result of our
analyses about the block orthogonal matching pursuit (BOMP) algorithm is
milder than those in the related work in \cite{re24}. Furthermore, we
characterize the data recovery condition from information theoretic point
of view.

\end{enumerate}

Thanks to the precoding operation and our sparsity-aware detection
algorithms, our scheme enable the system to support more active users to be
simultaneously served. The number of active users can be even more than the
number of antennas at BS. This is of great practical significance for
networks offering small packet services to a large number of users.

Applications of compressive sensing (CS) to random MAC channels have been
considered in \cite{re35}-\cite{re39}. In \cite{re35}, CS based decoding
scheme at the BS has been used for the multiuser detection task in
asynchronous random access channels. A technique based on CS for meter
reading in smart grid is proposed in \cite{re37}, and its consideration is
limited to single-antenna systems. Besides, a novel neighbor discovery
method in wireless networks with Reed-Muller Codes has been proposed in
\cite{re39}, where CS technique is also adopted. All the referred works
depend on the idea that the MAC channel is sparse, and all their works are
classified to initial category of CS, where no structure property have been
taken into account. This is one of the main distinctions that differentiate
our work from the referred ones.

The rest of the paper is organized as follows. In Section
\ref{SystemModel}, the system model of block-sparsity are given. In Section
\ref{sec.recovery}, we introduce the BOMP algorithm to recover the
transmitted signals, and discuss performance guarantees for data recovery.
Section \ref{simulations} will present the numerical experiments that
verify the effectiveness of our scheme.

\section{System Model}\label{SystemModel}

Assume the propagation environment is a block-fading channel and the
antennas at the BS, as well as the antennas among users, are uncorrelated
and uncoupled. We also assume that the transmissions are in blocks and the
users are synchronized at the block level. When a terminal has successfully
connected to the network, it becomes an online user, and the BS always has
the perfect channel state information (CSI) of online users. Our
consideration is only limited to uplink small packet transmission for
single-antenna users in massive MIMO system.

Consider an uplink system with $N$ mobile users, each with a single
antenna, and a base station with $M$ antennas. $N_a$ active users of the
total $N$ online users have small packets to send. For small packet
services, ${N_a} < N$, and usually ${N_a} \gg M$, even with massive MIMO,
we may have ${N_a} > M$. We assume that each frame of transmission consists
of $T$ symbols, and $T$ is no longer than the coherent interval of
block-fading channel. Let $\mathbf{s}_n \in {\mathbb{C}^{d \times 1}}$
denotes the symbols to be transmitted by user $n$, with $d < T$. User $n$
applies a precoding to $\mathbf{s}_n$ to yield
\begin{equation}
{\mathbf{x}_n} = {\mathbf{P}_n}{\mathbf{s}_n}
\end{equation}
where $\mathbf{P}_n$ is a complex precoding matrix of size $T\times d$. The
entries of $\mathbf{x}_n$ are transmitted in $T$ successive time slots. The
received signals at all antennas within one frame can be written as
\begin{equation}
\mathbf{Y} = \sqrt {{\rho _0}} \sum\limits_{n = 1}^N {{\mathbf{h}_n}\mathbf{x}_n^T}  + \mathbf{Z} =\sqrt {{\rho _0}} \sum\limits_{n = 1}^N {{\mathbf{h}_n}\mathbf{s}_n^T{\mathbf{P}_n^T}}  + \mathbf{Z}
\end{equation}
where ${\rho _0}$ is the signal to noise ratio (SNR) of the uplink,
$\mathbf{Y}$ is noisy measurement of size $M\times T$, $\mathbf{Z} \in
{\mathbb{C}^{M \times T}}$ represents the additive noise, with i.i.d.\
circularly symmetric complex Gaussian distributed random entries of zero
mean and unit variance, and $\mathbf{h}_n \in {\mathbb{C}^{M \times 1}}$
represents the channel coefficients from the user $n$ to the base station,
without loss of generally, let ${h_{mn}}\sim{\mathcal{CN}}\left( {0,1}
\right)$, $m = 1,2, \cdots ,M$. Using the linear algebra identity
$vec\left( {\mathbf{A}\mathbf{B}\mathbf{C}} \right) = \left(
{{\mathbf{C}^T} \otimes \mathbf{A}} \right)vec\left( \mathbf{B} \right)$,
where $vec$ denotes vectorizing $\mathbf{B}$ by column stacking and
$\otimes$ denotes the Kronecker product of two matrices, we can rewrite the
received signal as
\begin{equation} \label{model0}
vec(\mathbf{Y}) =\sqrt {{\rho _0}} \sum\limits_{n = 1}^N {(\mathbf{P}_n \otimes {\mathbf{h}_n}){\mathbf{s}_n}}  + vec(\mathbf{Z})
\end{equation}

Define $\mathbf{y}: = vec(\mathbf{Y})$, ${\mathbf{B}_n}: = \left(
{\mathbf{P}_n \otimes {\mathbf{h}_n}} \right)/\sqrt M $ and $\mathbf{B}: =
\left[ {{\mathbf{B}_1},{\mathbf{B}_2}, \cdots ,{\mathbf{B}_N}} \right]$,
$\mathbf{s}: = {\left[ {\mathbf{s}_1^T,\mathbf{s}_2^T, \cdots
,\mathbf{s}_N^T} \right]^T}$. Then we can write the model in (\ref{model0})
as
\begin{equation} \label{model1}
{\mathbf{y}} =\sqrt {{\rho _0}M} {\mathbf{B}}{\mathbf{s}}+ \mathbf{z}
\end{equation}

In this formulation, we have assumed that all the users have messages of
equal length $d$. This may not be the case in practice. We view $d$ as the
maximum length of the messages of all users within a frame. For the users
whose message length is less than $d$, we assume their messages have been
zero-padded to $d$ before precoding. Also, for those users that are not
active, we assume their transmitted symbols are all zeros.

Model (\ref{model1}) indicates that the signals to recover present the
structure of block-sparsity where transmitted signals are only located in a
small fraction of blocks and all other blocks are zeros. We collect all the
indices of blocks corresponding to active users to form a set $I$, with
$\left| I \right| = N_a$. When precoding matrix ${\mathbf{P}_n}$ is
reasonably designed, matrix ${\mathbf{B}}$ can meet the requirement for
sensing matrix in CS, and this kind of ${\mathbf{P}_n}$ is of wide range,
for instance, Gaussian or Bernoulli matrix. Therefore, model (\ref{model1})
can be viewed as block-sparse model in CS.

A few remarks about the precoding are needed. The precoding scheme is
proposed because in reality, $T$ is usually several times longer than the
lengths of small packets. Also, the precoding scheme contributes to solving
signal recovery problem in the situation where $N_a>M$. Secondly, each user
knows its own precoding matrix and the BS knows all precoding matrices of
all users. Thirdly, a basic requirement on the precoding matrix is that it
should be full column rank, which is a requirement for data recovery.
Additionally, in order to balance the power of every symbol of the messages
before and after being precoded, each column of ${\mathbf{P}_n}$ should be
normalized to unit energy. And finally, our precoding scheme is different
from spreading schemes in \cite{re38}, where direct sequence spread
spectrum (DSSS) is utilized for CS formulation.

\section{Data Recovery} \label{sec.recovery}

\subsection{BOMP Algorithm For Data Recovery}\label{BOMP}

The main idea of BOMP algorithm is that, for each iteration, it chooses a
block which has the maximum correlation with the residual signal, and after
that, it will use the selected blocks to approximate the original signals
by solving a least squares problem. In our scheme, we will adopt the BOMP
algorithm to recover the transmitted signal vector $\mathbf{s}$. About the
detailed calculation process of BOMP algorithm, we refer readers to article
\cite{re24}.

\subsection{Data Recovery Guarantees}\label{Guarantees}

In this section, we will present conditions that guarantee data recovery.
Before analyzing conditions for data recovery, some notation and
definitions will be introduced first. From the definition of
${\mathbf{B}}$, we can see that each column of it is statistically
normalized to one when number of antenna $M$ becomes large. Here we expand
${\mathbf{B}}$ as
\begin{equation}
{\mathbf{B}}   = \left[ {\underbrace {{\mathbf{b}_1} \cdots {\mathbf{b}_d}}_{\mathbf{B}_1}{\rm{ }}\underbrace {{\mathbf{b}_{d + 1}} \cdots {\mathbf{b}_{2d}}}_{\mathbf{B}_2} \cdots \underbrace {{\mathbf{b}_{\left( {N - 1} \right)d + 1}} \cdots {\mathbf{b}_{Nd}}}_{\mathbf{B}_N}} \right]
\end{equation}

As in \cite{re24}, we give the definitions of block-coherence in the form
of spectral norm
\begin{equation}
{\mu _\mathbf{B}}: = \frac{1}{d}\mathop {\max }\limits_{i \ne j} \left\| {\mathbf{B}_i^H}\mathbf{B}_j \right\|
\end{equation}
and sub-coherence as
\begin{equation}
\nu:  = \mathop {\max }\limits_{1 \le l \le N} \mathop {\max }\limits_{\left( {l - 1} \right)d + 1 \le i \ne j \le ld} \left| {\mathbf{b}_i^H{\mathbf{b}_j}} \right|
\end{equation}

At the same time define
\begin{equation}
{{s_l}} : = \mathop {\min }\limits_{{i \in I}} {\left\| {\mathbf{s}_i} \right\|_2}{\rm{ ~~~~    }} {{s_u}} : = \mathop {\max }\limits_{{i \in I}} {\left\| {\mathbf{s}_i } \right\|_2}
\end{equation}

In the following we will give two theorems to characterize conditions for
signal recovery, the proofs of which will be presented in the journal
version of this paper.

\subsection{Data Recovery Conditions For BOMP Algorithm} The following theorem
characterizes the block-sparse data recovery performance by BOMP algorithm.

\begin{theorem} \itshape
Consider the block-sparse model above, suppose that condition
\begin{equation}   \label{th_variables}
\begin{array}{l}
{\rho _0}M{\left[ {1 - \left( {d - 1} \right)\nu } \right]^2}{{{s_l^2}}} > {\tau ^2}\\
{\rm{   + }}{\rho _0}Md{\mu _\mathbf{B}}\left\{ {2\left( {{N_a} - 1} \right)\left[ {1 + \left( {d - 1} \right)\nu } \right] + {N_a^2}d{\mu _\mathbf{B}}} \right\}{{{s_l^2}}}\\
{\rm{   + }}2\sqrt {{\rho _0}M} \tau \left\{ {\left( {2{N_a} - 1} \right)d{\mu _\mathbf{B}} + \left[ {1 + \left( {d - 1} \right)\nu } \right]} \right\}{{s_l}}
\end{array}
\end{equation}
is satisfied, then the BOMP algorithm identifies the correct support of
signal vector $\mathbf{s}$ and at the same time achieves a bounded error
given by
\begin{equation}   \label{err_bound}
\left\| {\widehat {\mathbf{s}} - \mathbf{s}} \right\|_2^2 \le \frac{{K{\tau ^2}}}{{{\left[ {1 - \left( {d - 1} \right)\nu  - \left( {K - 1} \right)d{\mu _{\mathbf{B}}}} \right]^2}{\rho _0}M}}
\end{equation}
where $\widehat {\mathbf{s}}$ is the signal vector recovered by BOMP
algorithm, $K \le \left\lfloor {\frac{{MT}}{{{d}}}} \right\rfloor $ is the
maximum number of iterations for BOMP algorithm, $1 - \left( {d - 1}
\right)\nu - \left( {K - 1} \right)d{\mu _\mathbf{B}} > 0$ and $\tau =
\mathop {\max }\limits_{1 \le j \le N} {\left\| {\mathbf{B}_j^H\mathbf{z}}
\right\|_2}$.
\end{theorem}

\remark Since ${T} > d$ and $M{T} \gg d$, we can design orthogonal columns
for precoding matrix ${\mathbf{P} _n}$ of user $n$, $n = 1,2, \cdots ,N$,
then each block of matrix ${\mathbf{B}} $ is sub-matrix with orthogonal
columns, meaning $\nu = 0$. On the other hand, we have $\tau \gg {{s_l}} $
when each nonzero element of $\mathbf{s}_n$ satisfies a reasonable power
constrain. Additionally, if ${\mu _\mathbf{B}} = 0$, then condition
(\ref{th_variables}) can be simplified as ${{\rho _0}M}{{{s_l^2}}} > {\tau
^2} + 2\sqrt {{\rho _0}M}\tau {{s_l}} \approx {\tau ^2}$, which is milder
when compared with \cite[Theorem 5]{re24}, where result ${{\rho
_0}M}{{{s_l^2}}} > 4{\tau ^2}$ is given when applied to our scenario.

\remark In our scheme, when the number of active users are more than that
of the antennas at BS, the channel vectors among users are no longer
orthogonal or asymptotically orthogonal. However, by our precoding scheme,
correlations among columns in $\mathbf{B}$ can be smaller than correlations
among channel vectors of different users, which means that block-coherence
${\mu _\mathbf{B}}$ can still be rather small, as long as precoding is well
designed.

\subsection{Condition From Information Point Of View}
From the BS's point of view, it is desirable to recover all the information
conveyed by $\mathbf{s}$, including number of active users, exact indices
of these active users, their transmitted information bits, etc.. When all
the information are measured by bits, then The number of bits representing
the indices of active users and signal bits of the transmitted messages are
respectively $\log_2
\binom{N}{N_a}$ and $\sum_{i = 1}^{N_a} b_i$. Assume all bits are generated
with equal probability, and let $S$ denote the set of bits needed to
represent the total information, then we have $\left| S \right| \ge {\log
_2}{N \choose {N_a}} + \sum\nolimits_{i = 1}^{{N_a}} {{b_i}} $.

The following theorem characterizes the data recovery problem from
information theoretic point of view. Its proof is not included due to lack
of space.

\begin{theorem}\itshape
Define $p_e$ as the probability that some error has happened in the
recovery of information in set $S$, then the following condition is
necessary for the data recovery
\begin{equation}   \label{IT}
| S | \le  \frac 1 {1-p_e} [H(p_e)+\log_2\det (\mathbf{I}_{MT}+\rho
_0\mathbf{B}_I\mathbf{B}_I^H)]
\end{equation}
\end{theorem}

\section{Numerical Results}\label{simulations}

The experimental studies for verifying the proposed scheme are presented in
this section. In all simulations, the channel response matrix is i.i.d.
Gaussian matrix of complex values and the $N_a$ active users are chosen
uniformly at random among all $N$ online users. As for the block-sparse
data vectors to be transmitted, we assume quadrature phase shift keying
(QPSK) for data modulation. The symbol error rate (SER) and frame error
rate (FER) are used as the performance metrics. In all simulations, we do
not set the number of antennas to a large value, say one hundred or more,
for the sake of simplicity. We will choose the frame length to be a
multiple of the maximum length of short messages. We assume that all
messages have the same length $d$ unless otherwise specified. We simply
design $\mathbf{P}_n$ a random matrix with $(v = 0)$ or without $(v \ne 0)$
orthogonal columns, $n = 1,2, \ldots ,N$.

\begin{figure}[ht]
\centering
\includegraphics[width=\figwidth]{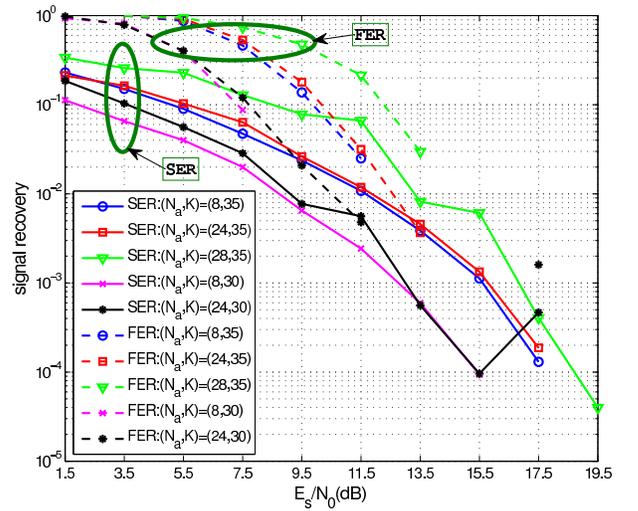}
\caption{signal recovery with $8$ antennas at BS}
\label{8_SER_FER1}
\end{figure}

\testcase Figure \ref{8_SER_FER1} shows the performance of the proposed
scheme with $8$ antennas at BS, where $K$ is the iterative number for BOMP
algorithm. Other parameters are given as $\left( {N,d,T,v} \right) = \left(
{80,200,1000,v = 0} \right)$. The results indicate that, the SER decreases
when ${E_s}/{N_0}$ increases and at the same time, increases when the
number of active users becomes larger. For case where iterative number is
$35$, when the number of active users is lower than a certain number, say
$24$ in our results, the SER is basically independent of this number. When
the number of active users exceeds the certain number, the performance will
get a lot worse; see e.g., ${N_a}=28$ in our results as an example.
Besides, the results we obtain for $8$ and $24$ active users are nearly
identical to those achieved by least square algorithm when active users are
already picked out and other $K - N_a$ off-support users are chosen at
random. Also in Figure \ref{8_SER_FER1}, we give results when less
iterations are employed for BOMP algorithm. When there are not too many
active users, less iterations benefit a lot, just like $30$ iterations for
$8$ active users. But for $24$ active users, fluctuations exist which means
$30$ iterations are not enough to include all the active users.

FER is also presented in Figure \ref{8_SER_FER1} with the same parameters
as that for SER. In our simulation, the FER is counted as follows: when
more than $8$ bits in a message are demodulated in error, we claim a frame
error, and if the bit errors are equal to or less than $8$, we hypothesize
that they can be detected and corrected by some channel coding schemes,
such as BCH Code. The same trend in performance of FER can be observed as
that of SER. As the ${E_s}/{N_0}$ increases, FER decreases quickly and when
the ${E_s}/{N_0}$ exceeds a certain value, the FER will be negligible.

The normalized throughput is defined as $\left( {1 - {P_{FER}}}
\right){N_a}d/\left( {M{T}} \right)$, where $\left( {1 - {P_{FER}}}
\right){N_a}$ is the maximum number of allowed active users in our scheme,
$P_{FER}$ is the value of FER; and $MT/d$ is the maximum number of users
that can be served when all timeslots of a frame are effectively used for
data transmission, which is $40$ under the given parameters. Take only the
signaling needed for resources competition before data transmission into
account, if $24$ active users are allowed to be simultaneously served,
since our scheme requires no additional signaling messages for data
transmission, the throughput will reach 60\%; while by conventional random
access protocols, if we regard the signaling messages (such as
request-to-send (RTS) signaling \cite{re6}) as some kinds of small packets
we considered, its throughput will be no more than 60\%, and if collision
happens, which is often the case, the throughput will decrease a step
further. Therefore, our scheme will greatly improve the system throughput.

\begin{figure}[ht]
\centering
\includegraphics[width=\figwidth]{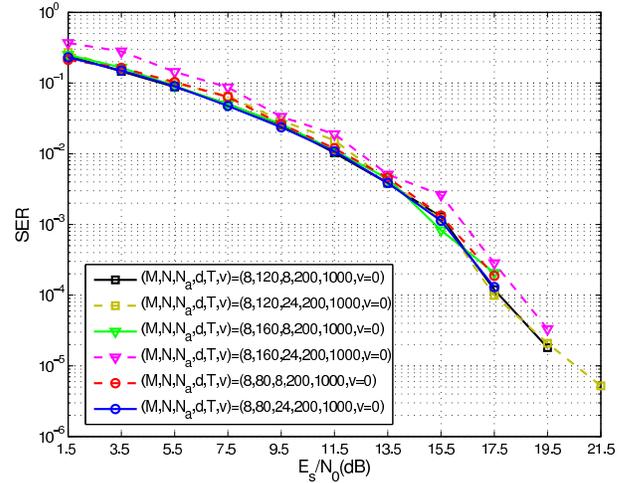}
\caption{signal recovery with different numbers of online users}
\label{potential_users3}
\end{figure}

\testcase By Figure~\ref{potential_users3}, we can see that when the number
of active users is fixed, the SER increases as the number of online users
increases, but the performance degradation is rather small, even when the
number of online users has been doubled, nearly no more than $1$dB
degradation can be observed for $24$ active users. By \textsl{Theorem 2},
the number of online users is not the dominant factor to affect the
performance under the given parameters.

\begin{figure}[ht]
\centering
\includegraphics[width=\figwidth]{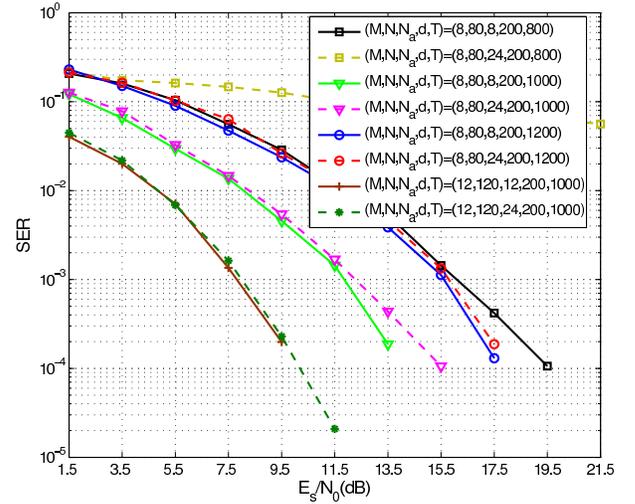}
\caption{symbol recovery with different numbers of BS antennas and with
different frame lengths}
\label{spreadT_antennas}
\end{figure}

\testcase Figure \ref{spreadT_antennas} depicts the performance when BS are
equipped with different numbers of antennas, and when frame lengths are
different. The results show that, under the same ratio ${N_a}/M$, when the
number of antennas $M$ increases, the SER performance becomes remarkably
better and a higher ratio ${N_a}/M$ can be accepted, which suggests a
higher throughput. With massive MIMO technique, this benefit can be reaped.
On the other hand, a big performance gap between antenna numbers of $8$ and
$12$ is observed, for the reason of iterative number demonstrated by Figure
\ref{8_SER_FER1}. More antennas at BS allows a larger iterative number for
BOMP algorithm to accommodate more active users, and the big performance
gap appears when we set both cases to the same number $35$ of iterations.
The curves respectively for $T=4d$, $T=5d$ and $T=6d$ show that, the longer
the frame length is, the better performance we can achieve, and thus the
more users that can be simultaneously served. However, affected by the
normalization of columns in precoding matrix, even when the length of frame
grows, the benefits diminish. This phenomenon will be observed when
parameters are chosen to ensure that $MT/(dK)$ is a constant.

\begin{figure}[ht]
\centering
\includegraphics[width=\figwidth]{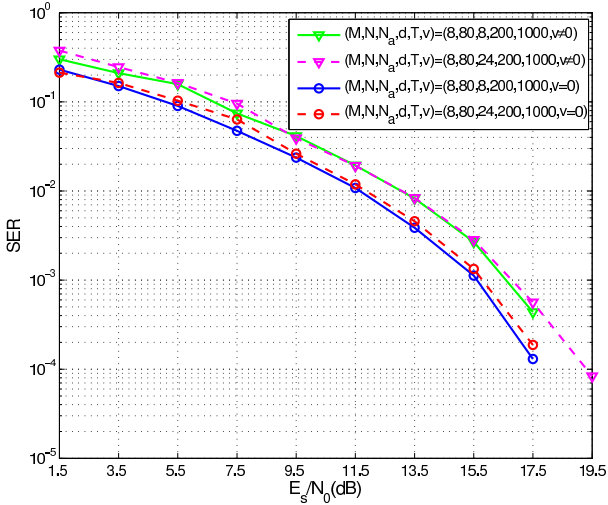}
\caption{symbol recovery with or without column orthogonal blocks}
\label{orthogonal}
\end{figure}

\testcase In Figure \ref{orthogonal}, we examine the performance when each
block of the matrix is column orthogonal or not. As we can see, for column
non-orthogonal case, nearly $1$dB performance degradation is observed when
compared with that of column orthogonal case, since the former case will
amplify the influence of noise to some degree.

\remark In all simulations, we set the number of antennas at BS just $8$ or
$12$, which is not enough for massive MIMO. In fact, the expected
performance of more antennas at BS can be significantly better, as
demonstrated by Figure \ref{spreadT_antennas}.

\remark In all simulations, we have set $d$ as the length of all messages.
In practice, this may not be the case. In fact, when different lengths for
messages exist and the number of active users is large, it has some
performance degradation. We have done the simulation and the result shows
that the performance gap is slight under wide conditions.

\remark The results given by our \textsl{Theorem 1} in terms of allow
active users is pessimistic whem compared with our simulation results. For
example, under the same conditions of Figure \ref{8_SER_FER1}, when
${E_s}/{N_0} = 10 dB$, only $1$ active user is allowed by
(\ref{th_variables}). This is because in the derivation of \textsl{Theorem
1}, we always consider the least favorable situation.

\section{Conclusion}\label{Conclusion}

In this paper, we propose an uplink data transmission scheme for small
packets. The proposed scheme combines the techniques of CS and massive
MIMO. Particularly, under the assumption that the BS has perfect CSI of
every online user and by a precoding scheme for block signal transmission,
we develope a block-sparse system model and adopte BOMP algorithm to
recover the transmitted data.

The transmission scheme addressed in this paper is applicable to future
wireless communication system. The reason is that small packet plays a more
and more important role in data traffic with the wide use of intelligent
terminals. The overall throughput of such a system is hampered by small
packet because of its heavy signaling overhead. Our scheme will greatly
reduce the signaling overhead and improve the throughput of the system.

\section*{Acknowledgment}
The work was partially supported by Chinese Ministry of Science and
Technology with grant No.2014AA01A704 and National Science Foundation of
China with grant No.61171112.

\end{document}